\documentclass[12pt]{article}
\usepackage{amssymb}
\usepackage{amsmath}
\usepackage{graphicx,color}

\usepackage{graphics}
\usepackage{color}
\usepackage{amssymb,amsmath}
\usepackage{slashed}
\usepackage{epstopdf}
\usepackage{booktabs}
\usepackage{mathrsfs}
\usepackage[export]{adjustbox}
\usepackage{epsfig}
\usepackage{graphicx,color}

\newcommand{\beq}{\begin{equation}}
\newcommand{\eeq}{\end{equation}}
\newcommand{\bea}{\begin{eqnarray}}
\newcommand{\eea}{\end{eqnarray}}
\newcommand{\beas}{\begin{eqnarray*}}
\newcommand{\eeas}{\end{eqnarray*}}
\newcommand{\bi}{\begin{itemize}}
\newcommand{\ei}{\end{itemize}}

\begin{document}
%
\begin{titlepage}
%
\vspace*{10mm}
\begin{center}
\baselineskip 25pt 
{\Large\bf
Enhanced pair production of heavy Majorana\\ neutrinos at LHC
}

\end{center}
\vspace{5mm}
\begin{center}
{\large
Arindam Das$^{~a,}$\footnote{arindam@kias.re.kr},
Nobuchika Okada$^{~b,}$\footnote{okadan@ua.edu}, 
and
Digesh Raut$^{~a,}$\footnote{draut@crimson.ua.edu}
}

\vspace{.5cm}

{\baselineskip 20pt \it
$^a$School of Physics, KIAS, Seoul 130-722, Korea\\
$^b$Department of Physics and Astronomy, University of Alabama, Tuscaloosa, AL35487, USA\\
} 

\end{center}
\vspace{0.5cm}
\begin{abstract}
Towards experimental confirmations of the type-I seesaw mechanism, we explore a prospect of discovering the heavy Majorana right-handed neutrinos (RHNs) from a resonant production of a new massive gauge boson ($Z^{\prime}$) and its subsequent decay into a pair of RHNs  ($Z^{\prime}\to NN$) at the future high luminosity runs at the Large Hadron Collider (LHC). 
Recent simulation studies have shown that the discovery of the RHNs through this process is promising in the future. 
However, the current LHC data very severely constrains the production cross section of the $Z^{\prime}$ boson into a dilepton final states, $pp \to Z^{\prime}\to \ell^{+}\ell^{-} $ ($\ell=e$ or $\mu$).  
Extrapolating the current bound to the future, 
we find that a significant enhancement of the branching ratio ${\rm BR}(Z^{\prime}\to NN$) over ${\rm BR}(Z^{\prime}\to \ell^{+}\ell^{-}$) is necessary for the future discovery of RHNs.    
As a well-motivated simple extension of the Standard Model (SM) to incorporate the $Z^\prime$ boson and the type-I seesaw mechanism,    we consider the minimal U(1)$_X$ model, which is a generalization of the well-known minimal $B-L$ model without extending the particle content. 
We point out that this model can yield a significant enhancement up to 
  ${\rm BR}(Z^{\prime}\to NN)/{\rm BR}(Z^{\prime}\to \ell^{+}\ell^{-}) \simeq 5$ (per generation). 
This is in sharp contrast with the minimal $B-L$ model, a benchmark scenario commonly used in simulation studies, which predicts ${\rm BR}(Z^{\prime}\to NN)/{\rm BR}(Z^{\prime}\to \ell^{+}\ell^{-}) \simeq 0.5$ (per generation). 
With such an enhancement and a realistic model-parameter choice to reproduce the neutrino oscillation data, 
 we conclude that the possibility of discovering RHNs with, for example, a $300 \; {\rm fb}^{-1}$ luminosity implies that 
  the $Z^\prime$ boson will be discovered with a luminosity of $170.5 \;{\rm fb}^{-1}$ ($125 \; {\rm fb}^{-1}$) 
  for the normal (inverted) hierarchy of the light neutrino mass pattern. 

\end{abstract}
\end{titlepage}

\setcounter{footnote}{0}


Although neutrinos are massless particles in the Standard Model (SM), the experimental evidence of the neutrino oscillation \cite{PDG} indicate that neutrinos have tiny but non-zero masses and flavor mixings. 
Hence, we need to extend the SM to incorporate the non-zero neutrino masses and flavor mixings.  
From a perspective of low energy effective theory, one can do so by introducing a dimension-5 operator \cite{Weinberg:1979sa} involving the Higgs and lepton doublets, which violates the lepton number by $\Delta L=2$ units. 
After the electroweak (EW) symmetry breaking, the neutrinos acquire tiny Majorana masses suppressed by the scale of the dimension-5 operator. 
In the context of a renormalizable theory, the dimension-5 operator is naturally generated by introducing heavy Majorana right-handed neutrinos (RHNs), which are singlet under the SM gauge group, and integrating them out. 
This is the so-called type-I seesaw mechanism \cite{seesaw1,seesaw2,seesaw3,seesaw4,seesaw5}.

If the RHNs have masses around 1 TeV or smaller, they can be produced at the  Large Hadron Collider (LHC) with a smoking-gun signature of a same-sign dilepton in the final state, which indicates a violation of the lepton number. 
Since the RHNs are singlet under the SM gauge group, they can be produced only through their mixings with the SM neutrinos. 
To reproduce the observed light neutrino mass scale, $m_\nu = \mathcal{O}(0.1)$ eV,  
  through the type-I seesaw mechanism with heavy neutrino masses at 1 TeV, 
  a natural value of the light-heavy neutrino mixing parameter is estimated to be $\mathcal{O} (10^{-6})$. 
Hence, the production of RHNs at the LHC with an observable rate is unlikely.\footnote{
In the general parametrization for the neutrino Dirac mass matrix \cite{Casas:2001sr}, this mixing parameter can be large. 
However, it turns out to be still small $\lesssim 0.01$ \cite{DO} in order to satisfy a variety of experimental constraints, 
  such as the neutrino oscillation data, the electroweak precision measurements and the lepton-flavor violating processes. 
}

In the simplest type-I seesaw scenario, the SM singlet RHNs are introduced only for the neutrino mass generation, and play no other important role in physics.  
One of more compelling scenarios, which incorporate the type-I seesaw mechanism, is the gauged $B-L$ extended SM \cite{mBL1, mBL2, mBL3, mBL4, mBL5, mBL6} . 
In this model the global U(1)$_{B-L}$ (baryon number minus lepton number) symmetry in the SM is gauged and the RHNs play the essential role to cancel the gauge and mixed-gravitational anomalies. 
Associated with the $B-L$ symmetry breaking, the RHNs acquire their Majorana masses, and the type-I seesaw mechanism is automatically implemented after the EW symmetry breaking. 
This model provides a new mechanism for the production of the RHNs at the LHC. 
Since the $B-L$ gauge boson ($Z^\prime$) couples with both the SM fermions and the RHNs, once the $Z^\prime$ boson is resonantly produced at the LHC, its subsequent decay produces a pair of RHNs. 
Then, the RHNs decay into the SM particles through the light-heavy neutrino mixings: 
  $N \to W^{\pm}  \ell^{\mp} $,  $Z \nu_{\ell}$, $Z {\overline \nu_\ell}$, $h \nu_{\ell} $, and  $h {\overline \nu_\ell} $.

Recently, in the context of the gauged $B-L$ models \cite{Kang:2015uoc, Yanagita:2017, Accomando:2017qcs}, the prospect of discovering RHNs in the future high luminosity runs at the LHC has been explored by simulation studies on a resonant $Z^\prime$ boson production and its decay into a pair of RHNs. 
In Refs.~\cite{Kang:2015uoc, Accomando:2017qcs}, the authors have considered the trilepton final states, 
  $Z^\prime \to NN \to  \ell^{\pm} \ell^{\mp}\ell^{\mp}$ $\nu_{\ell}$ $ j j $. 
For example, in Ref.~\cite{Accomando:2017qcs} the signal-to-background ratio of $S/{\sqrt B} \simeq 10$ has been obtained 
   at the LHC with a 300 fb$^{-1}$ luminosity, 
   for the production cross section, $\sigma (pp \to Z^\prime \to NN \to  \ell^{\pm} \ell^{\mp}\ell^{\mp} \nu_{\ell} jj) = 0.37$ fb ($\ell = e$ or $\mu$),
   with the $Z^\prime$ and RHN masses fixed as $m_{Z^\prime} = 4$ TeV and  $m_N = 400$ GeV, respectively. 
In Ref.~\cite{Yanagita:2017}, the authors have considered the final state with a same-sign dimuon and a boosted diboson, 
    $Z^\prime \to NN \to  \ell^{\pm} \ell^{\pm}$ $W^{\mp}W^{\mp}$. 
For fixed masses, $m_{Z^\prime} = 3$ TeV and  $m_N = m_{Z^\prime}/4$, 
   they have obtained a cross section $\sigma (pp \to Z^\prime \to NN \to  \mu^{\pm} \mu^{\pm} W^{\mp}W^{\mp}) \simeq 0.1$ fb 
   for a 5$\sigma$ discovery at the LHC with a 300 fb$^{-1}$ luminosity.

Since the RHNs are produced from the $Z^\prime$ boson decay, in exploring the future prospect of discovering the RHNs we need to consider the current LHC bound on the $Z^\prime$ boson production, which is already very severe.\footnote{
In Ref.~\cite{Yanagita:2017}, the authors have considered the $U(1)_{(B-L)_{3}}$ model \cite{Alonso:2017uky}, in which only the third generation fermions couple to the $Z^\prime$ boson. Hence, the current LHC bound on the $Z^\prime$ boson production is not applicable to the model, although their simulation results, which we employ in this letter, are model-independent.
}$^{,}$\footnote{
For a $Z^\prime$ boson mass of around TeV, 
  the constraints from the Electroweak Precision Measurements,  
  for example, from a $Z$-$Z^\prime$ mixing, is very weak as investigated in Ref.~\cite{Appelquist:2002mw}. 
We can also consider the LEP-2 bound on effective 4-Fermi interactions mediated by $Z^\prime$ boson. 
It has been shown in Ref.~\cite{Okada:2016tci} that the LHC constraints are more severe 
  than the LEP-2 bound for $m_{Z^\prime} \lesssim 5$ TeV. 
} 
The primary mode for the $Z^\prime$ boson search at the LHC is via the dilepton final states, 
  $ pp \to Z^\prime \to \ell^+ \ell^-$ ($\ell=e$ or $\mu$).  
The current upper bound on the $Z^\prime$ boson production cross section times its branching ratio 
  into a lepton pair ($e^+e^-$ and $\mu^+ \mu^-$ combined) is given by $\sigma (pp \to Z^\prime \to \ell^+ \ell^-) \lesssim 0.2$ fb,  
  for $m_{Z^{\prime}}\gtrsim 3$ TeV at the LHC Run-2 with 36.1 fb$^{-1}$ luminosity \cite{ATLAS_Z_Search}. 
Since the number of SM background events is very small for such a high $Z^\prime$ boson mass region, 
   we naively scale the current bound to a future bound as   
\bea
\sigma (pp \to Z^\prime \to \ell^+ \ell^-)  \lesssim 0.2 \ {\rm fb}\times \frac{36.1}{\cal L}, 
\label{ZBound} 
\eea
where ${\cal L}$ (in units of fb$^{-1}$) is a luminosity at the future LHC. 
Here, we have assumed the worst case scenario, namely, 
  there is still no indication of the $Z^\prime$ boson production in the future LHC data. 
For example, at the High-Luminosity LHC (${\cal L} =300$ fb $^{-1}$), the bound becomes 
$\sigma (pp \to Z^\prime \to \ell^+ \ell^-) \lesssim 2.4\times 10^{-2}$ fb. 
Note that this value is much smaller than the RHN production cross section of ${\cal O}(0.1)$ fb obtained in the simulation studies. 
Taking into account the branching ratios 
$NN \to  \ell^{\pm} \ell^{\mp}\ell^{\mp} \nu_{\ell} jj$ 
and $NN \to  \ell^{\pm} \ell^{\pm} W^{\mp}W^{\mp}$, 
 the original production cross section $\sigma (pp \to Z^\prime \to NN)$ must be rather large. 
Therefore, an enhancement of the branching ratio 
  ${\rm BR}(Z^\prime \to NN)$ over ${\rm BR}(Z^\prime \to \ell^+ \ell^-)$ 
  is crucial for the discovery of the RHNs in the future. 

In the worst case scenario with the $300$ fb$^{-1}$ luminosity, 
  we estimate an enhancement factor necessary to obtain 
   $\sigma (pp \to Z^\prime \to NN \to  \ell^{\pm} \ell^{\mp}\ell^{\mp} \nu_{\ell} jj)$, 
   $\sigma (pp \to Z^\prime \to NN \to  \mu^{\pm} \mu^{\pm}$ $W^{\mp}W^{\mp})= {\cal O}(0.1)$ fb,
   while $\sigma (pp \to Z^\prime \to \ell^+ \ell^-) \lesssim 2.4\times 10^{-2}$ fb. 
For $m_N \gg m_W = 80.4 $ GeV, $m_Z= 91.2$ GeV, and $m_h = 125.09$ GeV, we estimate the  branching ratios as 
  $\text{BR}(N\to W \ell) \simeq 0.5$ and $\text{BR}(N\to Z \nu) \simeq \text{BR}(N\to h \nu) \simeq 0.25$, 
   where we have considered one generation only. 
With $\text{BR}(W\to \ell \nu) \simeq 0.1$, $\text{BR}(W\to jj) \simeq 0.7$, $\text{BR}(Z \to \ell^+ \ell^-) \simeq 0.034$, $\text{BR}(Z \to \nu \nu) \simeq 0.2$, and $\text{BR}(Z \to jj) \simeq 0.7$, we estimate ${\rm BR}( N N \to \ell^+ \ell^-\ell^- \nu jj)={\rm BR}( N N \to \ell^- \ell^+\ell^+ \nu jj)\simeq 0.04$ and ${\rm BR}(N N \to \ell^\pm \ell^\pm W^\mp W^\mp) \simeq 0.125$. 
Hence, in order to obtain $\sigma (pp \to Z^\prime \to N N \to  \ell^{\pm} \ell^{\mp}\ell^{\mp}  \nu_{\ell} jj) \gtrsim 0.37$ fb \cite{Accomando:2017qcs} and $\sigma (pp \to Z^\prime \to NN \to  \ell^{\pm} \ell^{\pm}W^{\mp}W^{\mp} \gtrsim 0.1$ fb \cite{Yanagita:2017} 
we find $\sigma (pp \to Z^\prime \to N N) \gtrsim 4.62$ fb and $0.8$ fb, respectively. 
Hence, the enhancement factors we need are
\bea
 \frac{\text{BR}(Z^\prime \to N N)} {\text{BR}(Z^\prime \to \ell^+ \ell^-)} \gtrsim 192\;  \rm{and}\;  33.3, 
 \label{EFactor1}
\eea
respectively. 
Note that we only have $\frac{\text{BR}(Z^\prime \to NN)} {\text{BR}(Z^\prime \to \ell^+ \ell^-)} \simeq 0.5$ 
  in the minimal $B-L$ model.

\begin{table}[t]
\begin{center}
\begin{tabular}{c|ccc|c}
      &  SU(3)$_c$  & SU(2)$_L$ & U(1)$_Y$ & U(1)$_X$  \\ 
\hline
$q^{i}_{L}$ & {\bf 3 }    &  {\bf 2}         & $ 1/6$       & $(1/6) x_{H} + (1/3) x_{\Phi}$    \\
$u^{i}_{R}$ & {\bf 3 }    &  {\bf 1}         & $ 2/3$       & $(2/3) x_{H} + (1/3) x_{\Phi}$  \\
$d^{i}_{R}$ & {\bf 3 }    &  {\bf 1}         & $-1/3$       & $-(1/3) x_{H} + (1/3) x_{\Phi}$  \\
\hline
$\ell^{i}_{L}$ & {\bf 1 }    &  {\bf 2}         & $-1/2$       & $(-1/2) x_{H} - x_{\Phi}$  \\
$e^{i}_{R}$    & {\bf 1 }    &  {\bf 1}         & $-1$                   & $-x_{H} - x_{\Phi}$  \\
\hline
$H$            & {\bf 1 }    &  {\bf 2}         & $- 1/2$       & $(-1/2) x_{H}$  \\  
\hline
$N^{j}_{R}$    & {\bf 1 }    &  {\bf 1}         &$0$                    & $- x_{\Phi}$   \\

$\Phi$            & {\bf 1 }       &  {\bf 1}       &$ 0$                  & $ + 2x_{\Phi}$  \\ 
\hline
\end{tabular}
\end{center}
\caption{
 Particle content of  the minimal U(1)$_X$ model, 
 where $i,j=1, 2, 3$ are the generation indices.  
Without loss of generality, we fix $x_\Phi=1$. 
}
\label{tab1}
\end{table}  

In this paper we consider a simple extension of the SM, 
   which can yield a significant enhancement 
   for $\frac{{\rm BR}(Z^\prime \to NN)} {{\rm BR}(Z^\prime \to \ell^+ \ell^-)}$ as we will see in the following. 
This model is based on the gauge group, SU(3)$_c$$\times$SU(2)$_L$$\times$U(1)$_Y$$\times$U(1)$_X$, 
   where U(1)$_X$ is realized as a linear combination of the SM U(1)$_Y$ and U(1)$_{B-L}$ symmetry 
   (the so-called non-exotic U(1) extension of the SM \cite{Appelquist:2002mw}). 
The particle content of the model is listed in Table~\ref{tab1}. 
The structure of the model is the same as the minimal $B-L$ model except for the U(1)$_X$ charge assignment. 
In addition to the SM particle content, this model includes three generations of RHNs required 
  for the cancellation of the gauge and the mixed-gravitational anomalies, 
  a new Higgs field ($\Phi$) which breaks the U(1)$_X$ gauge symmetry, 
  and a  U(1)$_X$ gauge boson ($Z^\prime$).
The U(1)$_X$ charges are defined in terms of two real parameters $x_{H}$ and $x_{\Phi}$, 
  which are the U(1)$_X$ charges associated with $H$ and $\Phi$, respectively. 
In this model $x_{\Phi}$ always appears as a product with the U(1)$_{X}$ gauge coupling and 
  is not an independent free parameter, which we fix to be $x_{\Phi}=1$ throughout this letter. 
Hence, U(1)$_X$ charges of the particles are defined by a single free parameter $x_H$. 
Note that this model is identical to the minimal $B-L$ model in the limit of $x_{H}=0$.

The Yukawa sector of the SM is then extended to include
\bea
\mathcal{L} _{Y}\supset -\sum_{i, j=1}^{3} Y_{D}^{ij} \overline{\ell_{L}^{i}} H N_{R}^{j}-\frac{1}{2} \sum_{i=k}^{3} Y_{N}^{k} \Phi \overline{N_{R}^{k \ c}} N_{R}^{k}+ \rm{h. c.}, 
\label{U1XYukawa}
\eea 
where the first and second terms are the Dirac and Majorana Yukawa couplings. 
Here we use a diagonal basis for the Majorana Yukawa coupling without loss of generality.
After the U(1)$_X$ and the EW symmetry breakings, U(1)$_X$ gauge boson mass, the Majorana masses for the RHNs, and neutrino Dirac masses are generated:
\bea
 m_{Z^\prime} = g_X \sqrt{4 v_\Phi^2+  \frac{1}{4}x_H^2 v_h^2} \simeq 2 g_X v_\Phi ,  \; \;
   m_{N^i}=\frac{Y_N^i}{\sqrt{2}} v_\Phi, \; \; m_{D}^{ij}=\frac{Y_{D}^{ij}}{\sqrt{2}} v_h,
  \eea   
where $g_X$ is the U(1)$_X$ gauge coupling,  $v_\Phi$ is the $\Phi$ VEV, $v_h = 246$ GeV is the SM Higgs VEV, and we have used the LEP constraint \cite{Carena:2004xs, Heeck:2014zfa} ${v_\Phi}^2 \gg {v_h}^2$.

\begin{figure}[t]
\begin{center}
\includegraphics[width=0.6\textwidth, height=7cm]{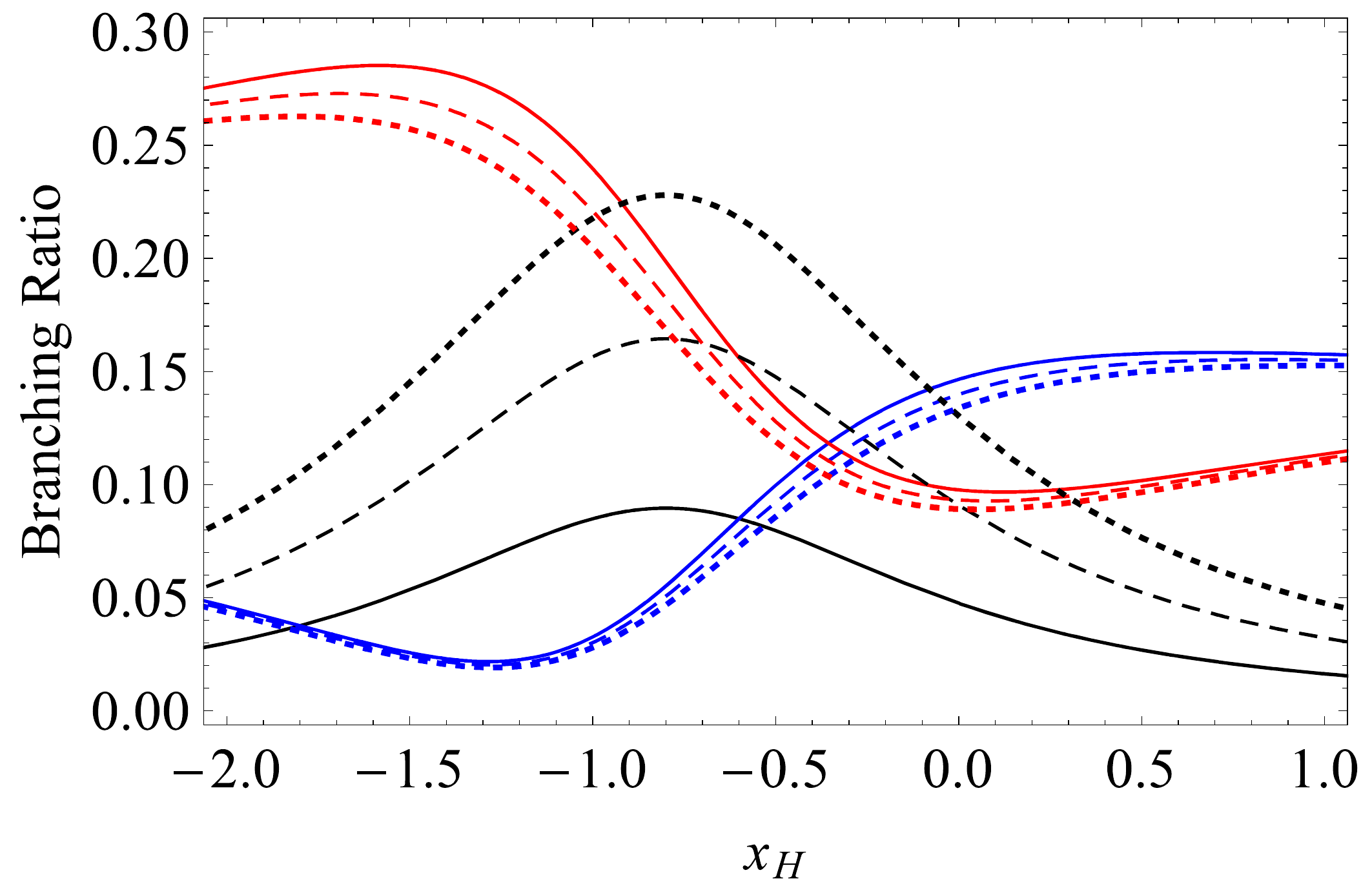} 
\caption{
The branching ratios of $Z^{\prime}$ boson 
   as a function of $x_H$ with a fixed $m_{ Z^{\prime}} = 3$ TeV. 
The solid lines correspond to $m_{N^1} = m_{ Z^{\prime}}/4$ and $m_{N^{2,3}} > m_{ Z^{\prime}}/2$;
the dashed (dotted) lines correspond to $m_{N^{1,2}} = m_{ Z^{\prime}}/4$ and $m_{N^{3}} > m_{ Z^{\prime}}/2$ ($m_{N^{1,2,3}} = m_{ Z^{\prime}}/4$ ). 
From top to bottom, the solid (red, black and blue) lines at $x_H = -1$ are the branching ratios to the first generations of jets (up and down quarks), 
  RHNs , and charged leptons, respectively. 
The lines for the RHN final states correspond to the sum of the branching ratio to all possible RHNs. 
}
\label{Branching1}
\end{center}
\end{figure}

Let us now consider the RHN production via $Z^{\prime}$ decay. 
The $Z^{\prime}$ boson partial decay widths 
 into a pair of SM chiral fermions ($f_L$) and a pair of the Majorana RHNs, 
 respectively, are given by 
\bea
\Gamma({Z^{\prime}\to {\overline{f_L}} f_L}) &=& N_c \; \frac{g_X^2 }{24 \pi} Q_{f_L}^2 m_{Z^{\prime}}, \nonumber \\
\Gamma(Z^{\prime}\to N^i N^i)
 &=& \frac{g^2 }{24\pi}  m_{Z^{\prime}} \left(1-\frac{4 m^2_{N^i}}{{m^2_{Z^\prime}}}\right)^{3/2}, 
\label{Zpwidths}
\eea 
where $N_c = 1 (3)$ is the color factor for lepton (quark), $ Q_{f_L}$ is the U(1)$_X$ charge of the SM fermion, 
  and we have neglected all the SM fermion masses. 
In Fig.~\ref{Branching1}, we show the $Z^{\prime}$ boson branching ratios for $m_{ Z^{\prime}} = 3$ TeV. 
The solid lines correspond to $m_{N^1} =m_{ Z^{\prime}}/4 $ and $m_{N^{2,3}} > m_{ Z^{\prime}}/2$, 
   the dashed (dotted) lines correspond to $m_{N^{1,2}} = m_{ Z^{\prime}}/4$ and 
   $m_{N^{3}} > m_{ Z^{\prime}}/2$ ($m_{N^{1,2,3}} = m_{ Z^{\prime}}/4$). 
For the SM final states, we show branching ratios to only the first generation dilepton and jets (sum of the jets from up and down quarks). 
The lines for the RHN final states correspond to the sum of the branching ratio to all possible RHNs. 
The plot shows the enhancement of RHNs branching ratios around $x_H = -0.8$  
  with the maximum values of the branching ratios, $0.09$, $0.16$, and $0.23$ 
  for the cases with one, two, and three generations of RHNs, respectively. 
For the minimal $B-L$ model ($x_H = 0$), the branching ratios are $0.05$, $0.09$, and $0.13$, respectively.

\begin{figure}[t]
\begin{center}
\includegraphics[width=0.6\textwidth, height=7cm]{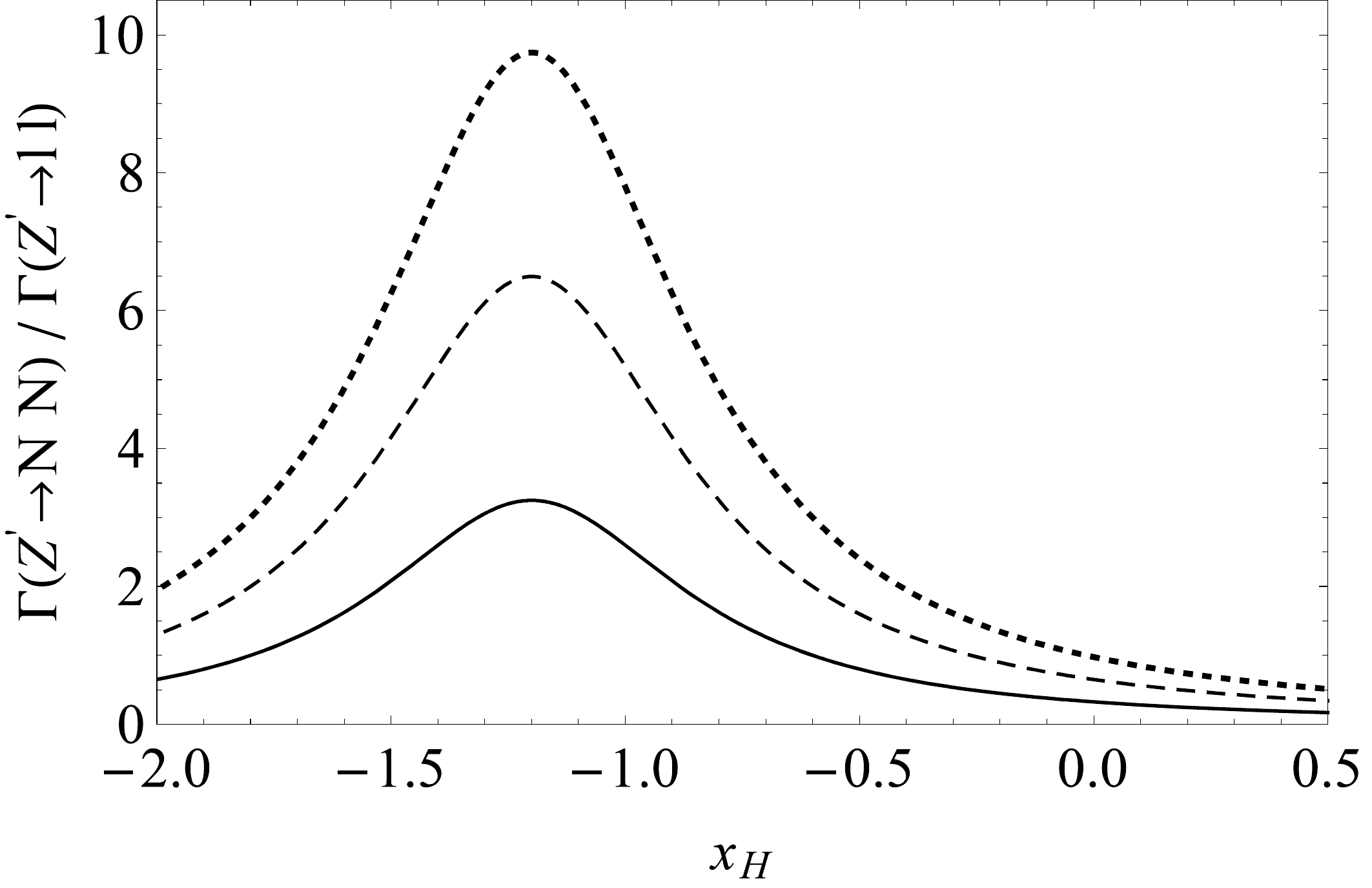}
\caption{The ratio of the partial decay widths of $Z^\prime$ boson into RHNs and dilepton final states as a function of $x_H$. 
The solid lines correspond to $m_{N^1} = m_{ Z^{\prime}}/4$ and $m_{N^{2,3}} > m_{ Z^{\prime}}/2$;
the dashed (dotted) lines correspond to $m_{N^{1,2}} = m_{ Z^{\prime}}/4$ and $m_{N^{3}} > m_{ Z^{\prime}}/2$ ($m_{N^{1,2,3}} = m_{ Z^{\prime}}/4$ ). }
\label{CSRatio}
\end{center}
\end{figure}

As we have discussed above, 
  the current LHC bound on the $Z^\prime$ boson production into the dilepton final states, 
  which is very severe, 
  requires $\frac{ {\rm BR}(Z^\prime \to NN)} {{\rm BR}(Z^\prime \to \ell^+ \ell^-)}\gg 1$ 
  for  the discovery of RHNs at the future LHC. 
This ratio is nothing but the ratio between the partial decay widths, 
  $\frac{\Gamma(Z^{\prime}\to N N)}{\Gamma({Z^{\prime}\to {\bar{\ell}} \ell})}$, 
  which is calculated from Eq.~(\ref{Zpwidths}) to be (per generation) 
\bea
\frac{\Gamma(Z^\prime \to NN)} {\Gamma(Z^\prime \to \ell^+ \ell^-)}
 = \frac{4}{8 + 12 x_H + 5 x_H^2} \left(1-\frac{4 m^2_N}{{m^2_{Z^\prime}}}\right)^{\frac{3}{2}}.  
\label{ZtoEE}
\eea 
With the same parameter choice as in Fig.~\ref{Branching1}, we show this ratio as a function of $x_H$ in Fig.~\ref{CSRatio}. 
We find the peaks at $x_H= -1.2$ with the maximum values of $3.25$, $6.50$, and $9.75$, respectively. 
Although we have obtained remarkable enhancement factors, 
  they do not reach the values required in the worst case scenario (see Eq.~(\ref{EFactor1})).  
Since the enhancement required for the trilepton final states is extremely large, 
  in the following we focus on the same sign dimuon and diboson final state, 
  which is the smoking-gun signature of the Majorana RHN production.
   
Let us now consider an optimistic case and assume that
the LHC experiment starts observing the $Z^\prime$ boson production through a dilepton final states with a luminosity below $300$ fb$^{-1}$. 
In this case we remove the constraint $\sigma (pp \to Z^\prime \to \ell^+ \ell^-) \lesssim 2.4\times 10^{-2}$ fb. 
Instead, we estimate the cross section $\sigma (pp \to Z^\prime \to \ell^+ \ell^-)$ 
  in order to achieve the RHN production cross section $\sigma (pp \to Z^\prime \to N N) \simeq 0.8$ fb 
  required for the $5\sigma$ discovery with the $300$ fb$^{-1}$ luminosity \cite{Yanagita:2017}. 
Let us fix $x_H = -1.2$ for which the ratio  ${\rm BR}(Z^{\prime}\to N N)/{\rm BR}(Z^{\prime}\to \ell^+ \ell^-)$ 
  reaches the maximum values of $3.25$, $6.50$, and $9.75$ for the cases with one, two, and three degenerate RHNs, respectively. 
Hence, we obtain $\sigma (pp \to Z^\prime \to \ell^+ \ell^-) \simeq 0.246$, $0.123$, and $0.0821$ fb for each case. 
The case with only one generation of RHN is already excluded by the current LHC results at 95 $\%$ confidence (see Eq.~(\ref{ZBound})). 
Since the number of SM background events is very small for a high $Z^\prime$ boson mass region ($m_{Z^\prime}\gtrsim 3$ TeV), let us here naively require 25 signal events for a 5-$\sigma$ discovery of the $Z^\prime$ boson production. 
Hence, the corresponding luminosities are found to be ${\cal L}({\rm fb}^{-1}) = 203$ and $305$ for the case with two and three RHNs, respectively. 
The required luminosities will be reached at the future LHC.

In the above analysis, we have simply assumed $\text{BR}(N\to W \mu) \simeq 0.5$. 
However, note that this branching ratio depends on the structure of the neutrino Dirac mass matrix,  
  and we expect $\text{BR}(N\to W \mu) < 0.5$ in a realistic parameter choice to reproduce the neutrino oscillation data. 
This implies that a more enhancement factor than what we have estimated above will be required 
  to obtain a sufficient number of signal events, while reproducing the neutrino oscillation data.

Let us look at the RHN decay processes in more detail. 
For simplicity, in the following analysis we consider the case with three degenerate RHNs. 
Assuming the hierarchy of $|m_D^{ij}/M_N| \ll 1$, we have the seesaw formula for the light Majorana neutrinos as 
\bea
m_{\nu} \simeq -\frac{1}{M_{N}} m_{D} m_{D}^{T}, 
\label{seesawI}
\eea
where $M_N = m_{N^{1}}= m_{N^{2}}= m_{N^{3}}$. 
We express the light neutrino flavor eigenstate $(\nu)$ in terms of the mass eigenstates 
  of the light $(\nu_m)$ and heavy $(N_m)$ Majorana neutrinos such as 
$\nu \simeq U_{\rm MNS} \; \nu_m+\mathcal{R} \; N_m$, 
  where $\mathcal{R} =m_D/M_N$, and $U_{\rm{MNS}}$ is the neutrino mixing matrix 
  by which $m_\nu$ is diagonalized as $U_{\rm MNS}^T m_\nu U_{\rm MNS}  = D_\nu = {\rm diag}(m_1, m_2, m_3)$.  
The heavy neutrino mass eigenstates have the charged current,  the neutral current, and 
  the Yukawa interactions as follows: 
\bea 
\mathcal{L}_{int} & \supset& 
 -\frac{g}{\sqrt{2}} W_{\mu}^+ \; 
  \overline{\ell_\alpha} \gamma^{\mu} P_L 
   {\cal R}_{\alpha j} N_m^j - \frac{g}{2 \cos \theta_{\rm W}}  Z_{\mu}  \; 
  \overline{\nu_\alpha} \gamma^{\mu} P_L {\cal R}_{\alpha j} N_m^j  - \frac{1}{\sqrt{2} v_h} h \; 
  \overline{\nu_{\alpha}} P_L {\cal R} _{\alpha j} N_m^j 
\label{Nint}  
\eea
where $\ell_\alpha$ and $\nu_\alpha$ ($\alpha=e, \mu, \tau$) are the three generations of 
  the charged leptons and neutrinos, 
  $P_L =  (1- \gamma_5)/2$,  and $\theta_{\rm W}$ is the weak mixing angle. 
Through the above interactions, a heavy neutrino mass eigenstate $N_m^i$ ($i =1,2,3$) decays 
  into $\ell_\alpha W$, $\nu_\alpha Z$, and $\nu_\alpha h$ with the corresponding partial decay widths: 
\bea
\Gamma(N_m^i \rightarrow \ell_{\alpha} W)
 &=& \frac{|R_{\alpha i}|^2}{16 \pi} 
 \frac{ (M_{N}^2 - m_W^2)^2 (M_{N}^2+2 m_W^2)}{M_{N}^3 v_h^2} , 
\nonumber \\
\Gamma(N_m^i \rightarrow \nu_{\ell_{\alpha}} Z)
 &=& \frac{|R_{\alpha i}|^2}{32 \pi} 
 \frac{ (M_{N}^2 - m_Z^2)^2 (M_{N}^2+2 m_Z^2)}{M_{N}^3 v_h^2} ,
\nonumber \\
\Gamma(N_m^i \rightarrow \nu_{\ell_{\alpha}} h)
 &=& \frac{|R_{\alpha i}|^2}{32 \pi}\frac{(M_{N}^2-m_h^2)^2}{M_{N} v_h^2} . 
\label{widths}
\eea

The elements of the matrix ${\cal R}$ are arranged to reproduce the neutrino oscillation data, 
 to which we adopt the following values: 
  $\sin^{2}2{\theta_{13}}=0.092$ \cite{DB} along with
 $\sin^2 2\theta_{12}=0.87$, $\sin^2 2\theta_{23}=1.0$, 
 $\Delta m_{12}^2 = m_2^2-m_1^2 = 7.6 \times 10^{-5}$ eV$^2$, 
 and $\Delta m_{23}^2= |m_3^2-m_2^2|=2.4 \times 10^{-3}$ eV$^2$ \cite{PDG}. 
Motivated by the recent measurement of the Dirac $CP$-phase, we set $\delta=\frac{3\pi}{2}$ \cite{CP-phase}, 
  while the Majorana phases are set to be zero for simplicity. 
From the seesaw formula we can generally parameterize the neutrino Dirac mass matrix as \cite{Casas:2001sr} 
\bea 
   m_D = \sqrt{M_N} U_{\rm{MNS}}^* \sqrt{D_\nu} \; O ,  
\label{mD}
\eea
where $\sqrt{D_\nu} \equiv {\rm diag}(\sqrt{m_1}, \sqrt{m_2}, \sqrt{m_3})$, and 
  $O$ is a general, complex $3\times3$ orthogonal matrix. 
With the inputs of the neutrino oscillation data and $M_N=m_{Z^\prime}/4$ with $m_{Z^\prime}=3$ TeV, 
  we have performed a parameter scan to find the maximum value of the branching ratio, 
  $\sum_{i=1}^3{\rm BR}(N_m^i  N_m^i \rightarrow \mu^\pm \mu^\pm W^\mp W^\mp)$. 
Here, for simplicity, we have considered $O$ to be a real orthogonal matrix, and fixed the lightest neutrino mass eigenvalue 
  to be $m_{\rm lightest} = 0.1 \times \sqrt{\Delta m_{12}^2}$. 
We have found the maximum values, 
  $\sum_{i=1}^3{\rm BR}(N_{i} N_{i} \rightarrow \mu^\pm \mu^\pm W^\mp W^\mp) \simeq  0.210$ ($0.154$), 
  for the normal (inverted) hierarchical light neutrino mass pattern.  
Using these realistic values, we now reconsider the optimistic case discussed above. 
For three degenerate RHNs, we previously obtained ${\cal L} = 305$ fb$^{-1}$ for a 5-$\sigma$ discovery of $Z^\prime$ boson production, which must be corrected to be ${\cal L}({\rm fb}^{-1}) = 170.5$  ($125$)  for the normal (inverted) hierarchy of the light neutrino mass pattern. 
Therefore, our scenario will be tested at the LHC in the near future.   
If we perform a general parameter scan for all free parameters, 
   the revised luminosity might become much larger. 
We leave the general parameter scan for future work \cite{FutureWork}.\footnote{
In our present analysis, we have considered the orthogonal matrix $O$ to be a real, for simplicity. 
In this case, a mixing between light and heavy neutrinos is of order $10^{-6}$, which is far below 
  the upper bounds form the electroweak precision measurements and the lepton flavor violating processes. 
See for example, Ref.~\cite{Das:2017nvm}.
}

In conclusion, we have investigated a prospect of discovering the RHNs in type-I seesaw at the LHC, 
  which are created from a resonant production of $Z^\prime$ boson and its subsequent decay into a pair of RHNs. 
Recent simulation studies have shown that the discovery of the RHNs is promising in the future. 
However, since the $Z^\prime$ boson generally couples with the SM charged leptons, 
  we need to consider the current LHC bound on the production cross section
  of the process, $pp \to Z^\prime \to \ell^+ \ell^-$ ($\ell=e$ or $\mu$), which is very severe. 
Under this circumstance, we have found that 
  a significant enhancement of ${\rm BR}(Z^{\prime}\to NN)/{\rm BR}(Z^{\prime}\to \ell^{+}\ell^{-})$  
  is necessary for the future discovery of the RHNs.  
As a simple extension of the SM, we have considered the minimal U(1)$_X$ model, 
  which is a generalization of the well-known minimal $B-L$ model. 
We have shown that this model can yield the significant enhancement of 
  $\frac{{\rm BR}(Z^{\prime}\to NN)}{{\rm BR}(Z^{\prime}\to \ell^{+}\ell^{-})} \simeq 3.25$ (per generation) 
  for $x_H = -1.2$, with $m_{Z^\prime} = 3$ TeV and $m_N = m_{Z^\prime}/ 4$. 
This is in sharp contrast with the minimal $B-L$ model, a benchmark model commonly used in simulation studies, 
  which predicts $\frac{{\rm BR}(Z^{\prime}\to NN)}{{\rm BR}(Z^{\prime}\to \ell^{+}\ell^{-})} \simeq 0.5$ (per generation). 
With this maximum enhancement factor and a realistic model-parameter choice to reproduce the neutrino oscillation data, we have concluded that the possibility of discovering RHNs with a $300 \; {\rm fb}^{-1}$ luminosity implies that the $Z^\prime$ boson 
 will be discovered with a luminosity of  $170.5 {\rm fb}^{-1}$ ($125 {\rm fb}^{-1}$)  for the normal (inverted) hierarchy of the light neutrino mass pattern.
When we employ $\sigma (pp \to Z^\prime \to NN \to  \mu^{\pm} \mu^{\pm} W^{\mp}W^{\mp}) \simeq 0.02$ fb 
   for the 5$\sigma$ discovery of RHNs with a 3000 fb$^{-1}$ luminosity \cite{Yanagita:2017}, 
   we simply scale, by a factor of 5, our results of the luminosity of $170.5 {\rm fb}^{-1}$ ($125 {\rm fb}^{-1}$)  
   for the $Z^\prime$ boson discovery to a luminosity of ${\cal L}({\rm fb}^{-1}) \simeq 853$ ($626$) 
   for the normal (inverted) hierarchical light neutrino mass pattern.  
From Eq.~(\ref{ZtoEE}), we can obtain an enhancement up to 
  $\frac{{\rm BR}(Z^{\prime}\to NN)}{{\rm BR}(Z^{\prime}\to \ell^{+}\ell^{-})} \simeq 5$ 
  if the mass splitting between the $m_N$ and $m_{Z^\prime}$ is larger, 
  which improves the prospect of discovering the RHNs in the future.

Finally, Fig.~\ref{Branching1} shows that 
  the $Z^\prime$ boson decay into $q \bar{q}$ final states is also enhanced at $x_H = -1.3$, 
  where we find 
  $\frac{\Gamma(Z^{\prime}\to q {\bar q})}{\Gamma(Z^{\prime}\to \ell^+ \ell^-)}=
  \frac{{\rm BR}(Z^{\prime}\to q {\bar q})}{{\rm BR}(Z^{\prime}\to \ell^+ \ell^-)} 
  =12.7$.  
One may think that with this enhancement the dijet final states could take the place of 
  the dilepton final states to become the primary search mode for the $Z^\prime$ boson production at the LHC. 
With this enhancement factor, the present bound on $\sigma (pp \to Z^\prime \to {\ell^+} \ell^-) \lesssim 0.2$ fb  
   is interpreted to the upper bound on $\sigma (pp \to Z^\prime \to \bar{q} q) \lesssim 2.54$ fb for $x_H = -1.3$.   
The recent result by the ATLAS collaboration with a 37 fb$^{-1}$ luminosity at the LHC Run-2 \cite{Aaboud:2017yvp} 
   has set the upper bound on $\sigma (pp \to Z^\prime \to \bar{q} q) \times A \lesssim 6$ fb 
   for $m_{Z^{\prime}}\simeq 3$ TeV, where $A < 1$ is the acceptance. 
Hence, the dilepton final states are still the primary search mode for the $Z^\prime$ boson production.

\section*{Acknowledgments}
This work of N.O. is supported in part by the U.S. Department of Energy (DE-SC0012447). 



\begin{thebibliography}{99}

\bibitem{PDG} 
C.~Patrignani {\it et al.} [Particle Data Group],
  ``Review of Particle Physics,''
  Chin.\ Phys.\ C {\bf 40}, no. 10, 100001 (2016).


\bibitem{Weinberg:1979sa} 
S.~Weinberg,
  ``Baryon and Lepton Nonconserving Processes,''
  Phys.\ Rev.\ Lett.\  {\bf 43}, 1566 (1979).


\bibitem{seesaw1} 
P.~Minkowski,
  ``$\mu \to e\gamma$ at a Rate of One Out of $10^{9}$ Muon Decays?,''
  Phys.\ Lett.\  {\bf 67B}, 421 (1977).
\bibitem{seesaw2}
T.~Yanagida,
  ``Horizontal Symmetry And Masses Of Neutrinos,''
  Conf.\ Proc.\ C {\bf 7902131}, 95 (1979).
%
\bibitem{seesaw3}
M.~Gell-Mann, P.~Ramond and R.~Slansky,
  ``Complex Spinors and Unified Theories,''
  Conf.\ Proc.\ C {\bf 790927}, 315 (1979)
  [arXiv:1306.4669 [hep-th]].
%
\bibitem{seesaw4}
S.~L.~Glashow, 
   ``Cargese Summer Institute: Quarks and Leptons," 
     Cargese, France, July 9-29, 1979, NATO Sci. Ser. B 61, 687 (1980).
%
\bibitem{seesaw5}
R.~N.~Mohapatra and G.~Senjanovic,
  ``Neutrino Mass and Spontaneous Parity Violation,''
  Phys.\ Rev.\ Lett.\  {\bf 44}, 912 (1980).



\bibitem{Casas:2001sr} 
  J.~A.~Casas and A.~Ibarra,
  ``Oscillating neutrinos and $\mu \to e~ \gamma$,''
  Nucl.\ Phys.\ B {\bf 618}, 171 (2001)
  [hep-ph/0103065].


\bibitem{DO}
A.~Das and N.~Okada,
  ``Bounds on heavy Majorana neutrinos in type-I seesaw and implications for collider searches,''
  Phys.\ Lett.\ B {\bf 774}, 32 (2017)
  [arXiv:1702.04668 [hep-ph]].


\bibitem{mBL1}
 R.~N.~Mohapatra and R.~E.~Marshak,
  ``Local B-L Symmetry of Electroweak Interactions, Majorana Neutrinos and Neutron Oscillations,''
  Phys.\ Rev.\ Lett.\  {\bf 44}, 1316 (1980)
  Erratum: [Phys.\ Rev.\ Lett.\  {\bf 44}, 1643 (1980)].

\bibitem{mBL2}
 R.~E.~Marshak and R.~N.~Mohapatra,
  ``Quark - Lepton Symmetry and B-L as the U(1) Generator of the Electroweak Symmetry Group,''
  Phys.\ Lett.\  {\bf 91B}, 222 (1980).

\bibitem{mBL3}
C.~Wetterich,
  ``Neutrino Masses and the Scale of B-L Violation,''
  Nucl.\ Phys.\ B {\bf 187}, 343 (1981).

\bibitem{mBL4}
A.~Masiero, J.~F.~Nieves and T.~Yanagida,
  ``$B^-$l Violating Proton Decay and Late Cosmological Baryon Production,''
  Phys.\ Lett.\  {\bf 116B}, 11 (1982).

\bibitem{mBL5}
R.~N.~Mohapatra and G.~Senjanovic,
  ``Spontaneous Breaking of Global $B-L$ Symmetry and Matter - Antimatter Oscillations in Grand Unified Theories,''
  Phys.\ Rev.\ D {\bf 27}, 254 (1983).


\bibitem{mBL6}
W.~Buchmuller, C.~Greub and P.~Minkowski,
  ``Neutrino masses, neutral vector bosons and the scale of B-L breaking,''
  Phys.\ Lett.\ B {\bf 267}, 395 (1991).




\bibitem{Kang:2015uoc} 
Z.~Kang, P.~Ko and J.~Li,
  ``New Avenues to Heavy Right-handed Neutrinos with Pair Production at Hadronic Colliders,''
  Phys.\ Rev.\ D {\bf 93}, no. 7, 075037 (2016)
  [arXiv:1512.08373 [hep-ph]]. 

  
\bibitem{Yanagita:2017} 
P.~Cox, C.~Han and T.~T.~Yanagida,
  ``LHC Search for Right-handed Neutrinos in $Z^\prime$ Models,''
  arXiv:1707.04532 [hep-ph].

 
\bibitem{Accomando:2017qcs} 
E.~Accomando, L.~Delle Rose, S.~Moretti, E.~Olaiya and C.~H.~Shepherd-Themistocleous,
  ``Extra Higgs Boson and $Z'$ as Portals to Signatures of Heavy Neutrinos at the LHC,''
  arXiv:1708.03650 [hep-ph]. 



\bibitem{Alonso:2017uky} 
  R.~Alonso, P.~Cox, C.~Han and T.~T.~Yanagida,
  ``Flavoured $B-L$ Local Symmetry and Anomalous Rare $B$ Decays,''
  arXiv:1705.03858 [hep-ph].
  

\bibitem{ATLAS_Z_Search} 
  M.~Aaboud {\it et al.} [ATLAS Collaboration],
  ``Search for new high-mass phenomena in the dilepton final state using 36 fb$^{-1}$ of proton-proton collision data at $ \sqrt{s}=13 $ TeV with the ATLAS detector,''
  JHEP {\bf 1710}, 182 (2017)
  [arXiv:1707.02424 [hep-ex]].



\bibitem{Appelquist:2002mw} 
  T.~Appelquist, B.~A.~Dobrescu and A.~R.~Hopper,
  ``Nonexotic neutral gauge bosons,''
  Phys.\ Rev.\ D {\bf 68}, 035012 (2003)
  [hep-ph/0212073].

\bibitem{Okada:2016tci} 
  N.~Okada and S.~Okada,
  ``$Z^\prime$-portal right-handed neutrino dark matter in the minimal U(1)$_X$ extended Standard Model,''
  Phys.\ Rev.\ D {\bf 95}, no. 3, 035025 (2017)
  [arXiv:1611.02672 [hep-ph]].




\bibitem{Carena:2004xs} 
  M.~Carena, A.~Daleo, B.~A.~Dobrescu and T.~M.~P.~Tait,
  ``$Z^\prime$ gauge bosons at the Tevatron,''
  Phys.\ Rev.\ D {\bf 70}, 093009 (2004)
  [hep-ph/0408098].


\bibitem{Heeck:2014zfa}
J.~Heeck,
  ``Unbroken B-L symmetry,''
  Phys.\ Lett.\ B {\bf 739}, 256 (2014)
  [arXiv:1408.6845 [hep-ph]].


\bibitem{DB}
  F.~P.~An {\it et al.} [Daya Bay Collaboration],
  ``Observation of electron-antineutrino disappearance at Daya Bay,''
  Phys.\ Rev.\ Lett.\  {\bf 108}, 171803 (2012)
  [arXiv:1203.1669 [hep-ex]].



\bibitem{CP-phase}
K.~Abe {\it et al.} [T2K Collaboration],
  ``Measurements of neutrino oscillation in appearance and disappearance channels by the T2K experiment with $6.6 \times 10^{20}$ protons on target,''
  Phys.\ Rev.\ D {\bf 91}, no. 7, 072010 (2015)
  [arXiv:1502.01550 [hep-ex]].

\bibitem{Das:2017nvm} 
  A.~Das and N.~Okada,
  ``Bounds on heavy Majorana neutrinos in type-I seesaw and implications for collider searches,''
  Phys.\ Lett.\ B {\bf 774}, 32 (2017)
  [arXiv:1702.04668 [hep-ph]].


\bibitem{FutureWork} 
  A. Das, N.~Okada, and D.~Raut, {\it work in progress}.  
  



\bibitem{Aaboud:2017yvp} 
 M.~Aaboud {\it et al.} [ATLAS Collaboration],
  ``Search for new phenomena in dijet events using 37 fb$^{-1}$ of $pp$ collision data collected at $\sqrt{s}=$13 TeV with the ATLAS detector,''
  Phys.\ Rev.\ D {\bf 96}, no. 5, 052004 (2017)
  [arXiv:1703.09127 [hep-ex]].


\end{thebibliography}
 \end{document}